\documentclass{ws-ijmpcs}

\begin{document}

\markboth{V. Bosch-Ramon}
{Non-Thermal Emission from Galactic Jets}

%
\catchline{}{}{}{}{}
%

\title{Non-Thermal Emission from Galactic Jets}

\author{Valent\'i Bosch-Ramon}

\address{Dublin Institute for Advanced Studies, Fitzwilliam Place 31\\
Dublin 2,
Ireland\footnote{}\\
valenti@cp.dias.ie}

\maketitle

\begin{history}
\received{Day Month Year}
\revised{Day Month Year}
\end{history}

\begin{abstract}
Jets are ubiquitous in the Universe. They are collimated outflows whose origin is associated to an accretion disc and a
central object, and can be very powerful non-thermal emitters. Jets form in active galactic nuclei, gamma-ray bursts,
microquasars, and young stellar objects. Galactic jets emitting non-thermal emission are typically associated to
microquasars, although the jets of massive young stellar objects are also non-thermal sources. The production of non-thermal
radiation, in particular radio synchrotron emission, is a clear indication that particle acceleration is taking place in the
source, which hints to the generation of photons even at high energies. In this work, we will discuss the emitting sites in,
or related to, microquasar jets, and briefly comment on the possibility of high-energy emission in jets from young stellar
objects.
\end{abstract}

\ccode{PACS numbers: 11.25.Hf, 123.1K}

\section{Introduction}

Microquasars are X-ray binaries with non-thermal jets (e.g. \refcite{mir99,rib05}). They are classified as high-mass microquasars when hosting a massive star, and low-mass microquasars
otherwise. Young stellar objects (YSO) are forming stars that can be low- or high-mass depending on the nature of the central object. YSO also produce jets, and some massive YSO present
non-thermal radio emission (e.g. \refcite{ara08}). 

The energy powering the non-thermal emission in microquasars can be either of accretion or black-hole rotation origin. The
magnetic and kinetic power is channelled through a jet launched from the inner regions of the accretion disk (e.g.
\refcite{bla77,bla82}), and part of this power is eventually converted into relativistic particles and radiation. In the case
of YSO, the jet, which supplies with energy the radiation processes, is powered by magneto-centrifugal processes or magnetic
pressure in the inner regions of an accretion disc. The energy may come from both the accretion disc, and the central object
rotation and magnetic fields (e.g. \refcite{rom09}; see also \refcite{vai11}).

For several decades, microquasars were considered strong candidates to gamma-ray sources (e.g. \refcite{cha85}; see also
\refcite{cha89,lev96,par00}), but they were not fully recognized as powerful gamma-ray emitters until recent years, after the
most recent generation of ground-based Cherenkov (HESS, MAGIC, VERITAS) and satellite-borne instruments ({\it Fermi}, {\it
AGILE}) arrived. The most relevant cases are the microquasars Cygnus~X-1\footnote{This source has been detected in GeV and
TeV energies with significances close, but slightly below, 5~$\sigma$, and thus these detections are still to be firmly
established.} and Cygnus~X-3 (\refcite{alb07,sab10,tav09,abd09a,sab11}). There are four other binary systems that may be also
gamma-ray emitting microquasars: LS~I~+61~303\cite{alb06,abd09b,pit09}, LS~5039\cite{aha05,abd09c,pit09},
HESS~J0632$+$057\cite{hin09,fal11,mol11}, and 1FGL~J1018.6$-$5856\cite{cor11}, although they could as well
host a non-accreting pulsar. Regarding the microquasar and pulsar scenarios, LS~I~+61~303 and LS~5039 have been extensively
discussed in the literature (see, e.g., \refcite{bos09} and references therein).

The detection of non-thermal radio emission in YSO associated sources like those of Serpens\cite{rod89}, or HH~80$-$81\cite{mar93}, was a clear indication that particle acceleration takes
place in these regions, probably at the termination of the jets\cite{cru90,hen91}. Interestingly, the jets themselves also show non-thermal radio emission\cite{reid95,car10}, which shows
that dissipation and subsequent particle acceleration already take place well before the jets are terminated in the ISM. 

Although the detected radio emission is already evidence of particle acceleration in microquasar jets, the finding of
microquasar gamma-ray emission proves that these sources can very efficiently channel accretion or black-hole rotational
energy into radiation. This high efficiency, together with the temporal characteristics of the detected radiation may favor
leptonic models, although hadronic mechanisms cannot be discarded. Also, the extreme conditions under which gamma-rays are
produced can put restrictions in the emitter structure. Morphological studies can be also of help, since non-thermal
processes can take place not only at the binary scales, but also far away (e.g. the jet termination region). Although the
complexity of microquasar phenomenology can make the characterization of the ongoing processes difficult, high quality data
together with semi-analytical modeling can provide sensible information on the non-thermal physics of the sources. Numerical
calculations are also important, since they can inform about the conditions of the background plasma in which emission is
taking place. 

In the case of YSO, the evidence of particle acceleration suggests that high-energy emission may be also taking place, and it
is necessary to explore this possibility, since the new and forthcoming gamma-ray instrumentation may provide information on
the physical processes taking place in these objects (see, e.g., \refcite{ara07,bos10}). In fact, there is statistical
evidence that YSO may be associated to some GeV sources, detected by the {\it Fermi} satellite and still
unidentified\cite{mun11}.

In this paper, we review relevant aspects of the non-thermal emission in microquasars. We focus mainly in the GeV and TeV energy bands, for which photon production requires extreme
conditions in these sources. At the end of this work, the high-energy emission in YSO is also briefly discussed.

\section{Non-thermal emission in microquasars}

Microquasar jets can produce non-thermal populations of relativistic particles via diffusive shock acceleration or other
mechanisms at different spatial scales. These particles, electrons, protons or even heavy nuclei, can interact with the
background matter, radiation and magnetic fields to produce non-thermal emission, sometimes from radio to gamma-rays. In the
GeV and TeV bands, the most efficient process is inverse Compton (IC), but in systems with very high density regions, like
SS~433, Cygnus~X-3 and possibly Cygnus~X-1, proton-proton interactions may be also relevant. Also, in systems with very dense
fields of energetic target photons, photomeson production and even photodisintegration of nuclei may be efficient. The
gamma-ray emission can take place at different scales, although certain regions suffer from strong absorption via pair
creation (e.g. deep inside the system or at the jet base), and some others may lack enough targets (e.g. the jet largest
scales). Below, we discuss farther the non-thermal phenomena at different scales in high- and low-mass microquasars. For a
general review on the efficiency of leptonic and hadronic processes under typical microquasar conditions, see \refcite{bos09}
and references therein. 

\begin{figure}[pb]
\centerline{\psfig{file=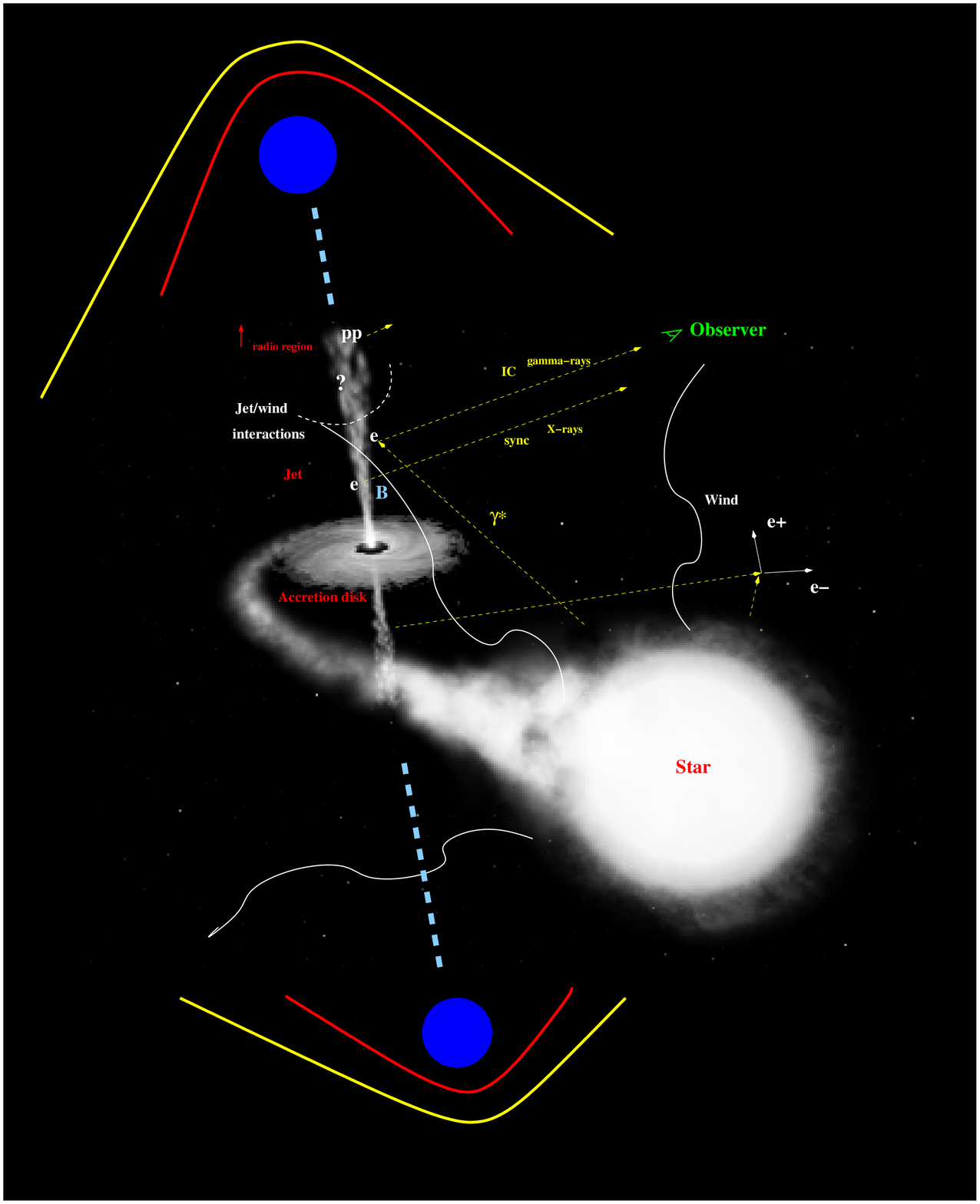,angle=0,width=9cm}}
\vspace*{8pt}
\caption{Illustrative picture of the microquasar scenario (not to scale), 
in which relevant elements and dynamical and radiative processes in different regions, 
including the jet termination regions, are shown (background image from ESA, NASA, and F\'elix Mirabel).\label{mq}}
\end{figure}

In Figure~\ref{mq}, a sketch of the microquasar
scenario is presented.

\subsection{Microquasar emitting sites}

Different emitting regions can be considered when understanding the non-thermal emission from microquasars. Jets are the best acceleration sites given the large amount of energy that they
transport. Different forms of dissipation can take place in them through shocks, velocity gradients and turbulence (e.g. \refcite{rie07}), as well as magnetic reconnection (e.g.
\refcite{zen01}), which can lead to generate different types of non-thermal particle populations. 

\begin{figure}[pb]
\centerline{\psfig{file=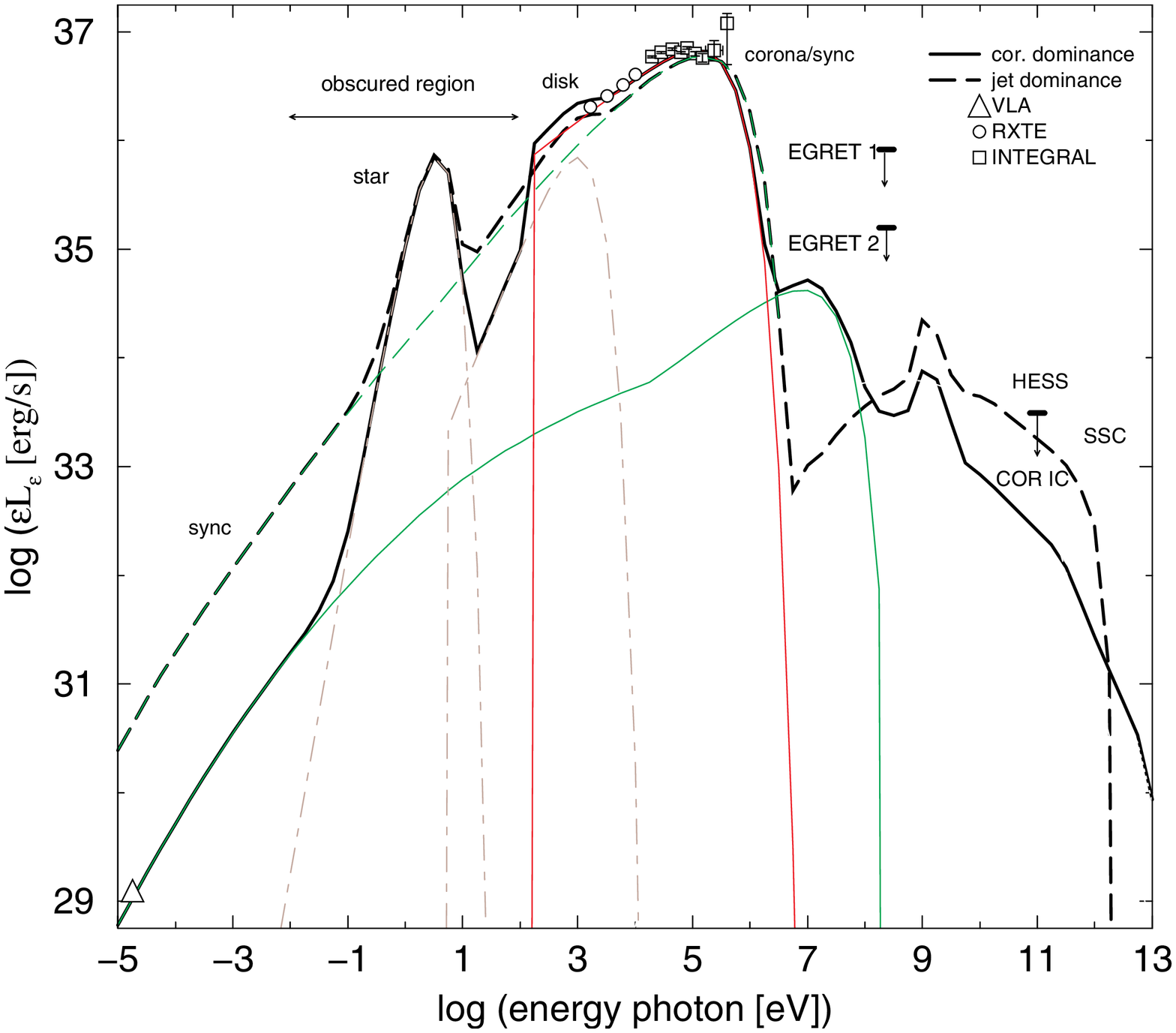,angle=270,width=9cm}}
\vspace*{8pt}
\caption{Computed spectral energy distribution of the non-thermal emission from 1E~1740.7$-$2942 for two situations. In one case, 
the hard X-rays come from a corona, whereas in the other, they are of synchrotron origin and come from the
jet. For details, see 38.\label{lmqs}}
\end{figure}

The jet formation itself, interaction with an accretion disc wind, or recollimation and internal shocks can accelerate
particles at the jet base. The presence of non-thermal electrons in the region can lead to the production of gamma rays
through IC scattering of accretion photons, or with photons produced by the same electrons via synchrotron emission (e.g.
\refcite{bos06b}). The base of the jet is possibly the region in which hadronic processes may be the most efficient, given
the high density of matter and photons in there, and the hardness of the latter (e.g. \refcite{lev01,rom08}). The local
radiation fields could also strongly suppress the GeV emission via gamma-ray absorption and pair creation (see, e.g.,
\refcite{rom08,cer11}). For low ambient magnetic fields, electromagnetic cascades can increase the effective transparency of
the source to gamma-rays\cite{akh85}. An example of a (leptonic) low-mass microquasar spectral energy distribution, with
its high-energy radiation mainly coming from the base of the jet, is shown in Fig.~\ref{lmqs}.

\begin{figure}[pb]
\centerline{\psfig{file=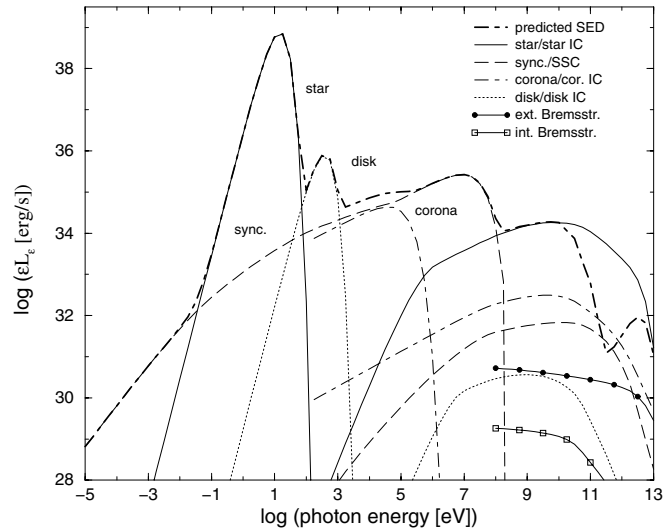,angle=270,width=9cm}}
\vspace*{8pt}
\caption{Computed spectral energy distribution, from radio to very high energies, for a high-mass microquasar (see 43).\label{hm}}
\end{figure}

In high-mass microquasars, at the scales of the binary system, the strong radiation and mass loss from the star can render
significant non-thermal radiation, in particular at high energies; radio emission may be at least partially free-free
absorbed. The considered most efficient high-energy channel is typically IC with stellar photons (e.g. \refcite{bos06a}),
interaction that is anisotropic and yields specific lightcurve and spectral features (e.g. \refcite{kha08}). For instance,
anisotropic IC may be behind the orbital modulation of the GeV lightcurve of Cygnus~X-3 seen by {\it
Fermi}\cite{abd09a,dub10}. Absorption of TeV emission in the stellar photon field is likely to be significant for compact
high-mass systems, like Cygnus~X-3 and Cygnus~X-1. That may be the reason why the former has not been detected in the TeV
range (see \refcite{ale10} and references therein), and why the evidence of detection of Cygnus~X-1 by MAGIC may imply an
emitter outside the binary system (see \refcite{bos08a}). As before, for low enough magnetic fields (see \refcite{kha08}),
electromagnetic cascades can increase the effective transparency of these two sources (see, e.g.,
\refcite{bed07,ore07,bed10}). As discussed below, the role of pair creation cannot be neglected in the context of broadband
non-thermal emission. At the binary scales, absorption of GeV photons is not expected since this band is below the gamma-ray
energy threshold for pair creation, around $\sim 10-100$~GeV for stellar photons peaking in the UV. Proton-proton, photomeson
production and photodisintegration have also been proposed as possible mechanisms of gamma-ray emission at these scales (e.g.
\refcite{rom03,aha06,bed05,rey08}). An example of a (leptonic) spectral energy distribution of a high-mass microquasar is shown in
Fig.~\ref{hm}.

\begin{figure}[pb]
\centerline{\psfig{file=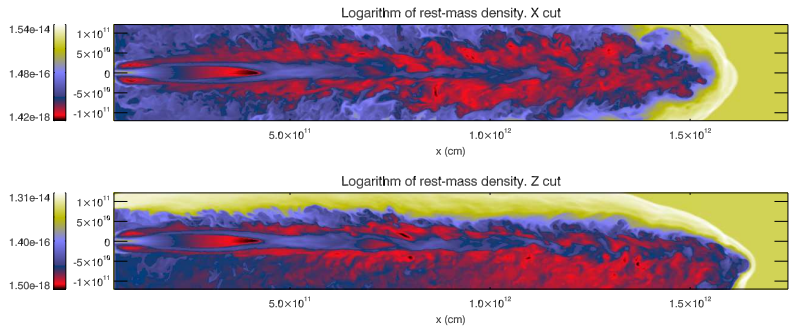,angle=0,width=12cm}}
\vspace*{8pt}
\caption{Density map for a high-mass microquasar jet interacting with the stellar wind, which is coming from the top of the 
image in the bottom figure, and perpendicular to the page in the top one (see 56).\label{jetint}}
\end{figure}

The interaction of the jets with the stellar wind, at the binary scales, cannot be neglected in microquasars with a massive
companion. The impact of the wind on the jet triggers strong shocks, good candidates for particle acceleration, jet bending,
and potentially jet disruption (e.g. \refcite{per08,per10}). This interaction can generate high-energy emission\cite{per08},
but the specific properties can depend on the level of inhomogeneity of the stellar wind (e.g. \refcite{ara09,ara11}; also
Perucho \& Bosch-Ramon, in preparation). Figure~\ref{jetint} shows the density map resulting from a 3-dimensional
hydrodynamical simulation of a microquasar jet interacting with the wind of the companion.

\begin{figure}[pb]
\centerline{\psfig{file=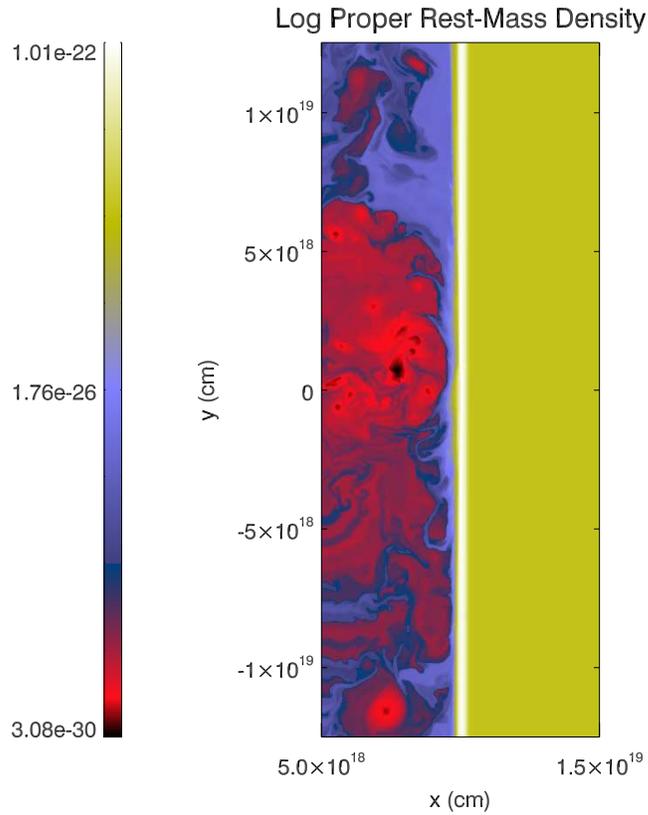,angle=0,width=9cm}}
\vspace*{8pt}
\caption{Density map resulting from a 2-dimensional slab simulation, in which
a jet propagates in an environment characterized by the microquasar motion in the ISM. The shocked stellar wind, deflected by the microquasar 
proper motion, comes from the top and impact the jet from a side. The jet is clearly destroyed by the medium 
interaction (see 61). \label{ls}}
\end{figure}

Far from the binary system, say at milliarcsecond to arcsecond scales, the jet propagates unaffected by significant external
disturbances. However, there are different mechanisms that may lead to energy dissipation, particle heating/acceleration and
subsequent radiation, like velocity gradients and turbulence triggered by Kelvin-Helmholtz instabilities in the jet walls.
Shear acceleration has been proposed for instance to explain extended emission from large scale jets in microquasars and
AGN\cite{rie07}. All this could generate fresh relativistic particles that could emit in radio by synchrotron. Very powerful
ejections could also be bright enough to be detectable, from radio to gamma-rays, far away from the binary (e.g.
\refcite{ato99}). It is noteworthy that, unless there is not significant previous jet activity, the wall of a continuous jet
or a transient ejection are always to encounter diluted and hot jet material. This material was reprocessed in the jet
reverse shock, where jet and ISM pressures balance, and was swept backwards filling the so-called cocoon. Only the presence
of a strong wind, either from the accretion disc or the star, can clean this material out up to a certain distance from the
microquasar. However, the jet material will unavoidably end up embedded in the cocoon plasma before the reverse shock is
reached. The pressure of the cocoon can trigger a recollimation shock in the jet, which becomes collimated and suffers
pinching. The jet fed cocoon drives a slow forward shock in the ISM, much denser and cooler than the jet. This complex
dynamical behavior has associated the production of non-thermal emission, which possibly may reach gamma-ray energies. An
interesting situation arises when the microquasar has a high-mass companion and the proper motion velocity is $\gtrsim
10^7$~cm~s$^{-1}$, in which case the jet can be completely disrupted before reaching the ISM, as illustrated in
Fig.~\ref{ls}. Farther discussion of jets interacting with the ISM can be found in \refcite{bor09}, \refcite{bos11b}, and
references therein.

\begin{figure*}
\centering
\includegraphics[angle=0, width=0.45\textwidth]{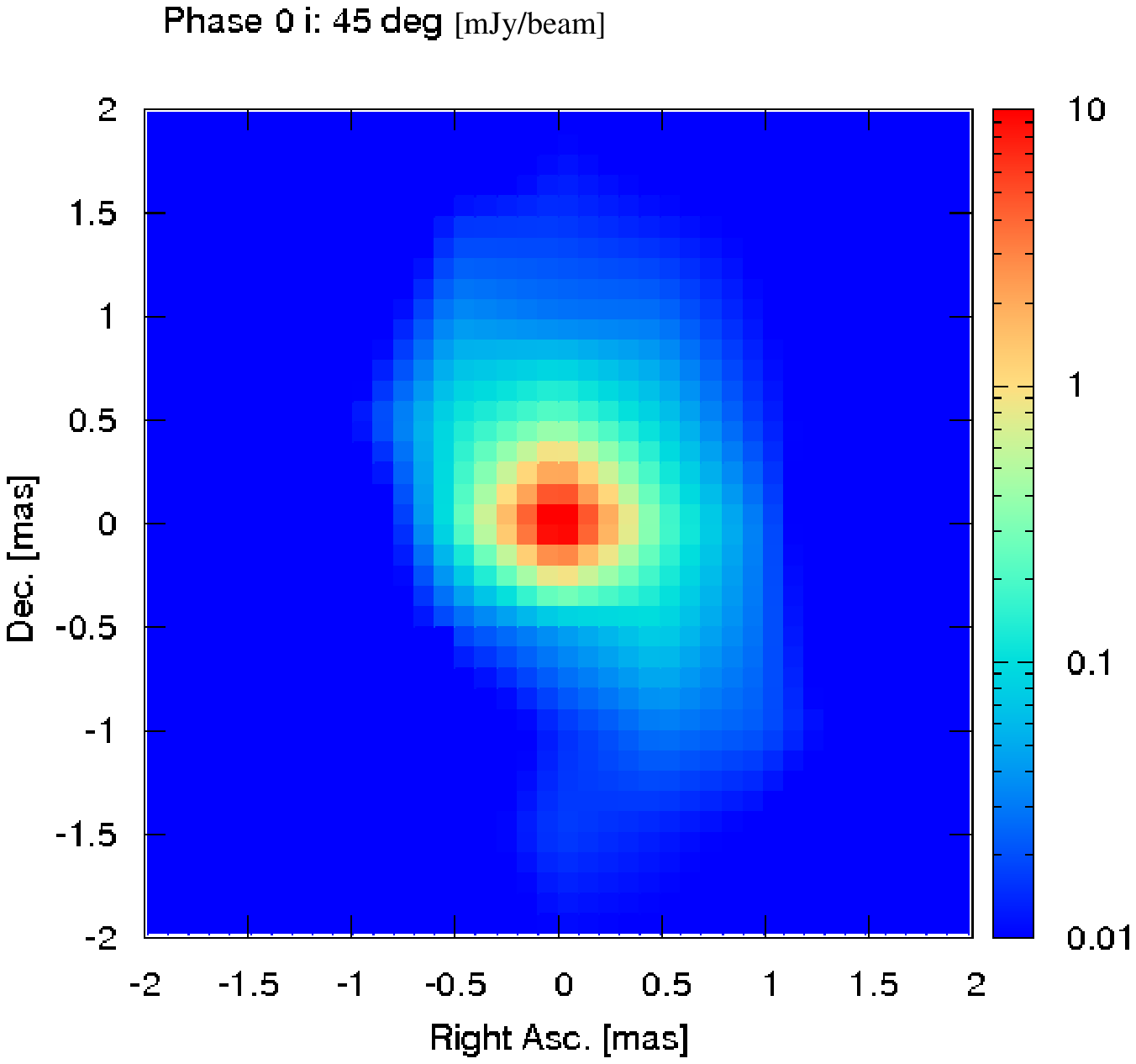}\qquad
\includegraphics[angle=0, width=0.45\textwidth]{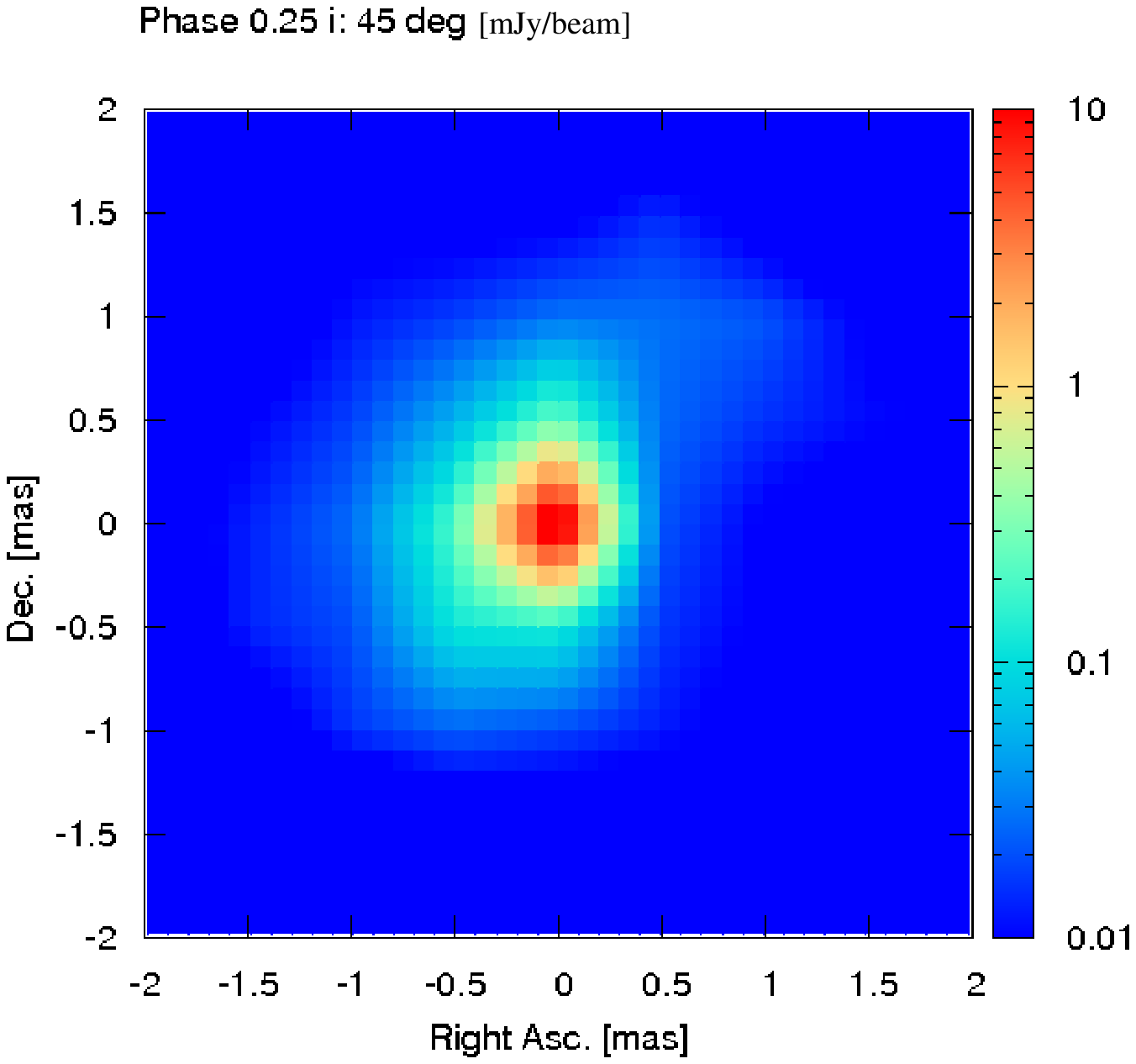}\\[10pt]
\includegraphics[angle=0, width=0.45\textwidth]{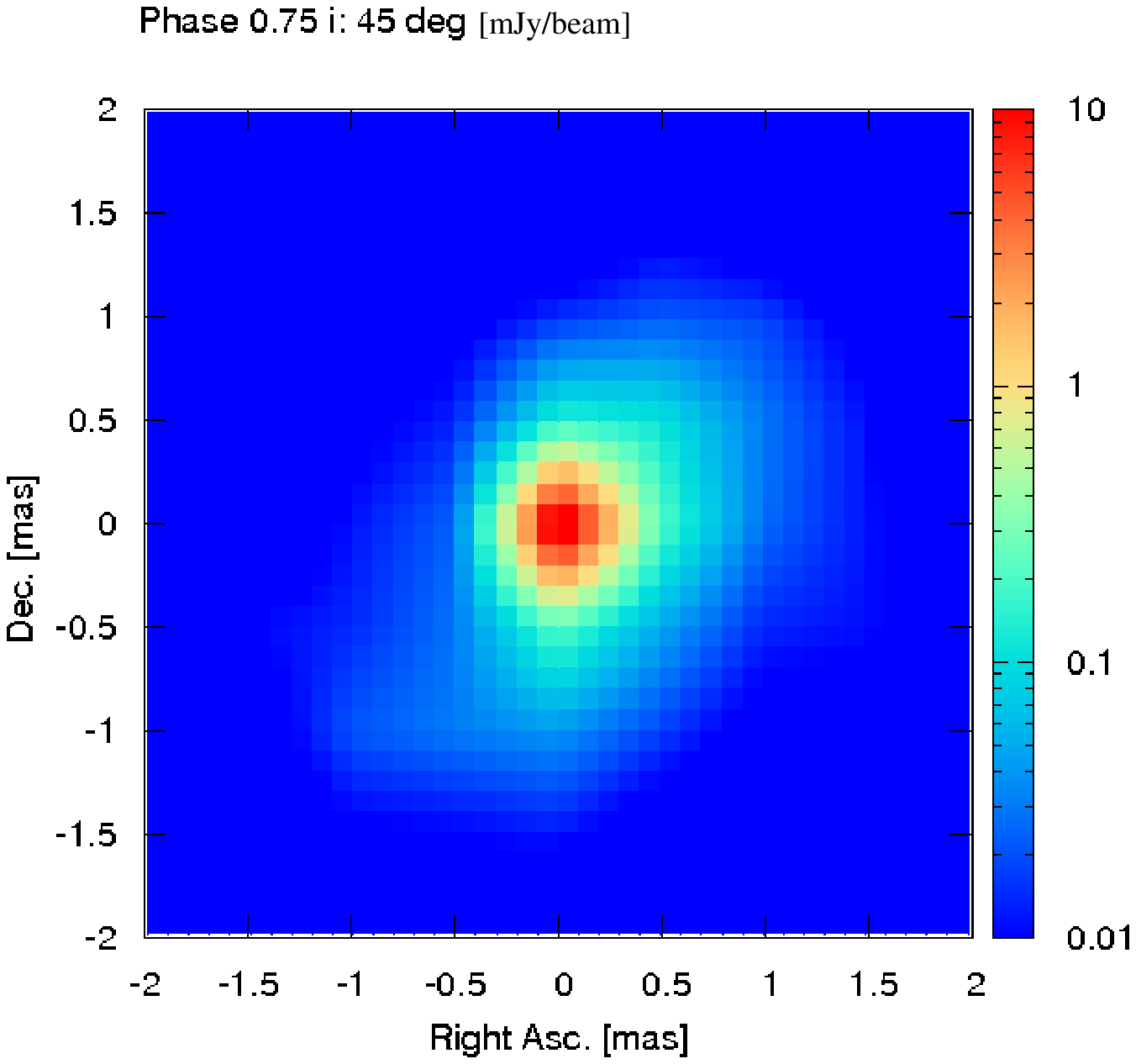}\qquad
\includegraphics[angle=0, width=0.45\textwidth]{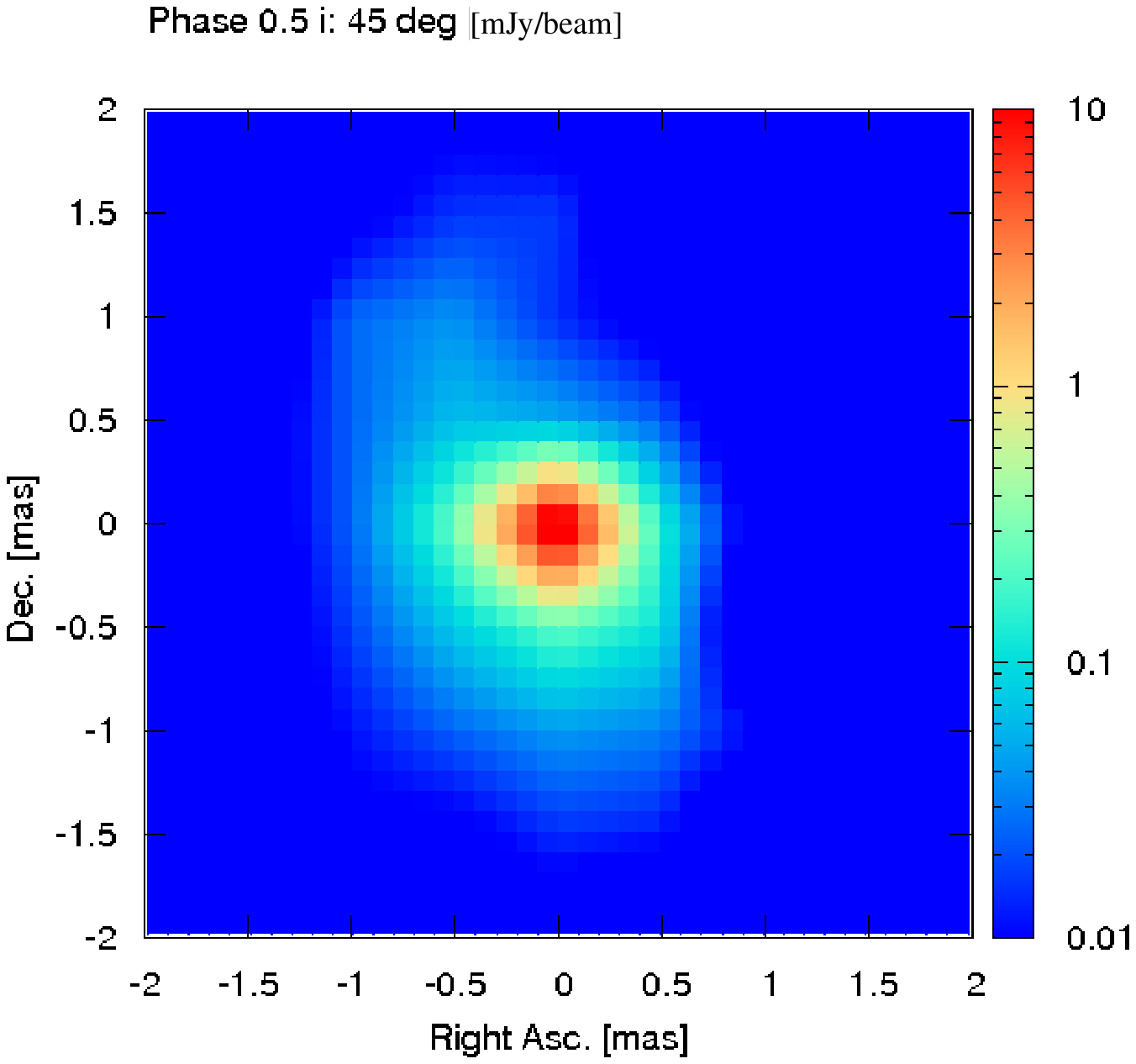}\\
\caption{Computed image, in the direction to the observer, of the 5~GHz radio emission from secondary pairs in a TeV binary for different orbital phases.
Units are given in mJy per beam, being the beam size $\sim 1$~milliarcsecond (see 68).}
\label{raw}
\end{figure*}

Jets, or their termination region, are not the only possible emitting sites in microquasars. The inner regions of the accretion structures (e.g. disc, corona/ADAF and the like) may also
contain non-thermal populations of particles (e.g. \refcite{bis76,pin82,spr88,gie99,rom10}). At the binary system scales, and in particular with high-mass companions, the stellar wind is
dense, carries magnetic field, and is embedded in a dense photon bath by the star. Therefore, for those systems with very efficient particle acceleration in the jet, electrons and protons
could diffuse out of it and radiate their energy in the environment. Also, gamma-ray absorption due to pair creation in the stellar photon field can inject electrons and positrons in the
wind, also leading to broadband non-thermal emission, as shown for instance in \refcite{bos08b}. This emission may be actually behind a substantial fraction of the milliarcsecond radiation in
a TeV emitting microquasar\cite{bos11a}. An example of this is shown in Fig.~\ref{raw}, in which 5~GHz maps are presented for different orbital phases in a TeV emitting binary.

\section{Young stellar objects at high energies}

\begin{figure}[pb]
\centerline{\psfig{file=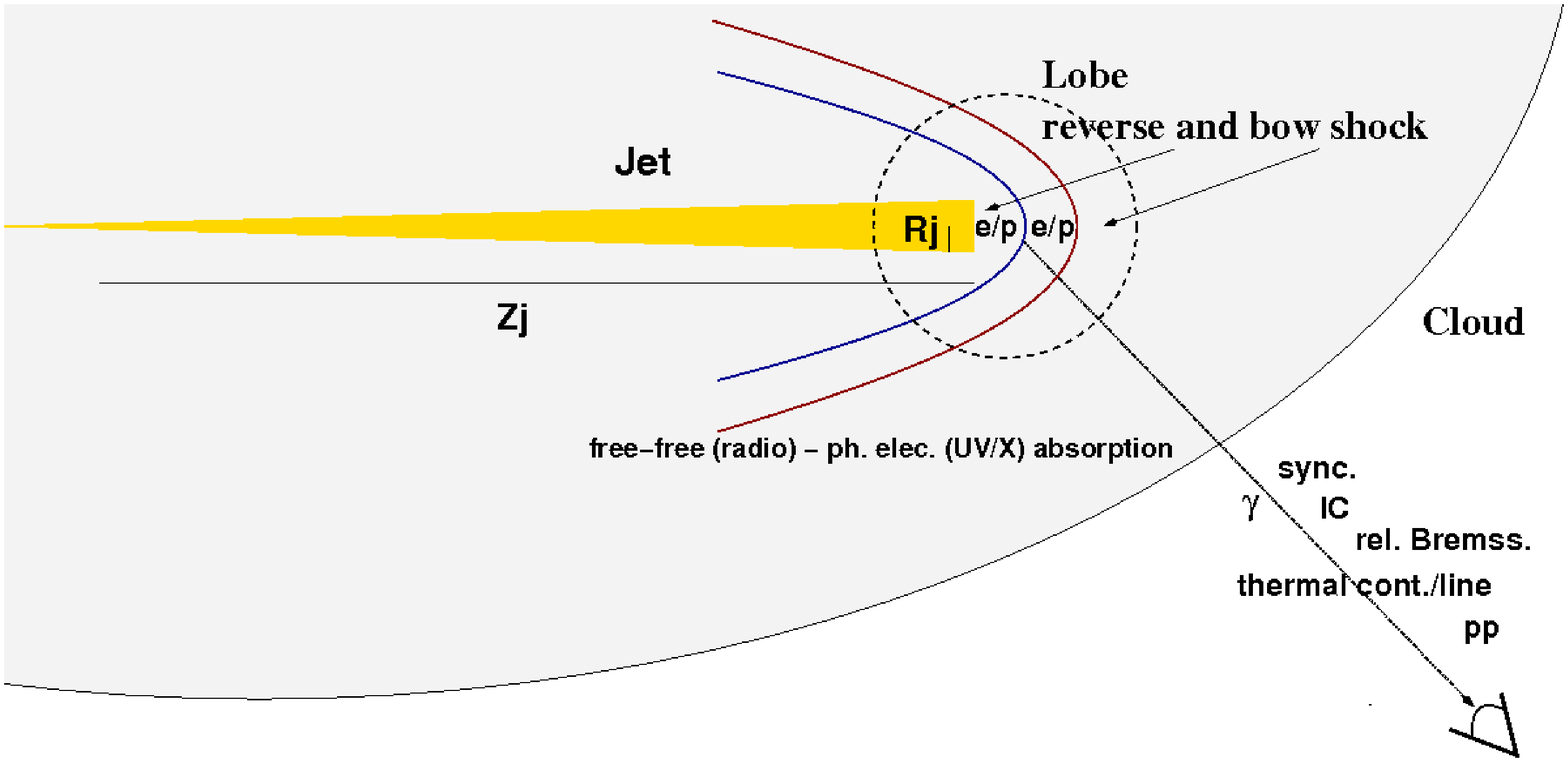,angle=0,width=11cm}}
\vspace*{8pt}
\caption{Sketch of the termination region of the jet of a massive YSO. Two shocks of different strengths and velocities 
(a reverse shock in the jet and a forward shock in the ambient medium) 
will form in the jet head depending on the jet-medium properties. Electrons and protons can be accelerated in the shocks, and generate 
nonthermal emission via interaction with the ambient matter, magnetic, and radiation fields. 
The shocked material will also produce thermal radiation.\label{skter}}
\end{figure}

\begin{figure}[pb]
\centerline{\psfig{file=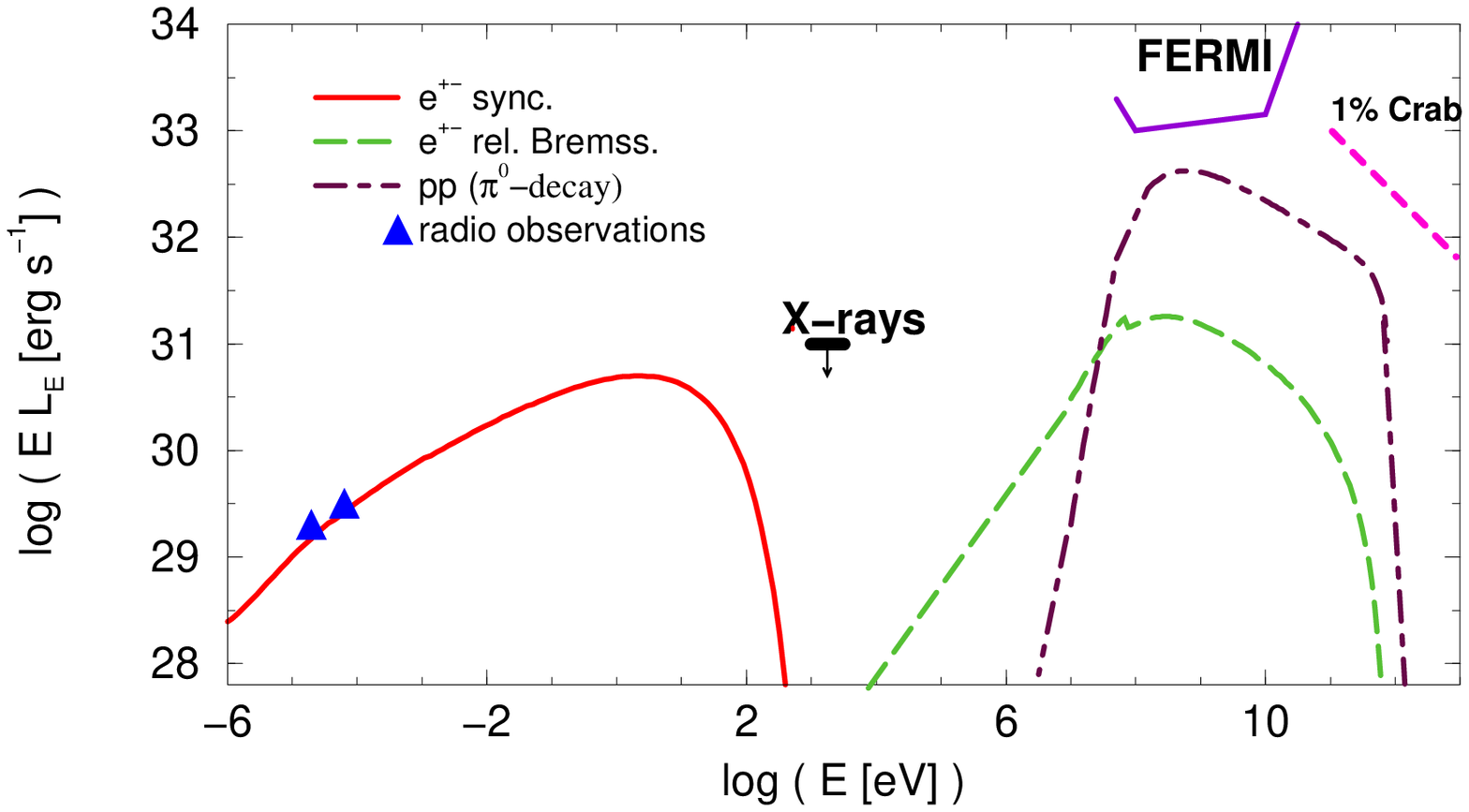,angle=0,width=11cm}}
\vspace*{8pt}
\caption{Spectral energy distribution of the nonthermal emission from one of the components of HH~80$-$81. 
The IC contribution is negligible and not shown here. 
Observational points are from IRAS~16547$-$4247 (radio, 67;
X-rays, 32). The 1~yr/5~$\sigma$ (optimistic) sensitivity of {\it Fermi} in the direction of the galactic 
plane is shown. A curve above 100~GeV showing a luminosity corresponding to 0.01~Crab, 
typical sensitivity of a Cerenkov telescope for exposures of $\sim 50$~hr, is also presented (see 33).\label{sedyso}}
\end{figure}

Recent work has shown that the termination region of jets in massive young stellar objects may be high-energy emitters (e.g.
\refcite{ara07,ara08,bos10}). The emission, which may reach hundreds of GeV, depending on the jet velocity and the local
magnetic field, would be produced through relativistic Bremsstrahlung and, in case protons were also accelerated, via
proton-proton collisions. The dominance of these processes with respect to IC stems from the ambient densities, very high
since the sources are massive forming stars embedded in dense molecular clouds, and the moderate ambient radiation fields. A
sketch of the scenario of high-energy emission from the termination region of a massive YSO jet is presented in
Fig.~\ref{skter}. A computed spectral energy distribution is shown in Fig.~\ref{sedyso} for illustrative purposes.

\section{Final remarks}

Microquasar can efficiently accelerate particles up to very high energies and produce gamma-rays within and outside the binary system. It is however unclear currently why some sources emit
gamma-rays and others do not. A key point may be the presence of a massive star, which as discussed here can affect the jet significantly, with the formation of particle acceleration sites,
and also offering dense target photon and matter fields, suitable for gamma-ray production at the binary scales. Low-mass microquasars could in principle produce gamma-rays, and a reason
for remaining undetected yet may be that their GeV emission is too dim for the present instrumentation. At TeV energies, in the context of leptonic models, strong synchrotron cooling and IC
scattering deep in the Klein-Nishina regime (as expected in the hard accretion photon fields of low-mass microquasars) may also prevent their detection in TeV. The lack of a dense stellar
wind in low-mass systems makes also a difference with high-mass ones. It is worth noting that all the gamma-ray microquasars or microquasar candidates harbor massive stars, which seems to
be a common feature in most of the known gamma-ray binaries (not only microquasars). At the jet base, the situation seems to be basically the same in high- and low-mass systems. In both
object types, the compactness of the region could imply that gamma rays are absorbed. Presently, gamma rays have not been detected at the largest scales, and thus it is not clear whether
there is an intrinsic difference between high- and low-mass microquasars at these scales, which may be the case accounting for their different environments. In the case of YSO, the local
conditions and the energetics may be suitable for gamma-ray emission, although specific combinations of magnetic fields and non-thermal powers are needed. In any case, a detection in GeV or
$\sim 100$~GeV in the near future seems feasible, and forthcoming instruments operating above few tens of GeV, like CTA, may indeed be able to see these sources.


\section*{Acknowledgments}
I want to thank the organizers for kindly invitating me to give this talk.
The research leading to these results has received funding from the European
Union
Seventh Framework Program (FP7/2007-2013) under grant agreement
PIEF-GA-2009-252463.
V.B.-R. acknowledges support by the Spanish Ministerio de Ciencia e 
Innovación (MICINN) under grants AYA2010-21782-C03-01 and 
FPA2010-22056-C06-02.






\end{document}